\documentclass[conference]{IEEEtran}

\ifCLASSINFOpdf
  \usepackage[pdftex]{graphicx}
  \graphicspath{
  {./}
  {figures/}
  {../figures/}
  }  
  
 \DeclareGraphicsExtensions{.pdf,.jpeg,.png}
\else
\fi

\usepackage{epstopdf}
\usepackage[cmex10]{amsmath}
\usepackage{array}
\usepackage{mdwmath}
\usepackage{mdwtab}
\usepackage[caption=false,font=footnotesize]{subfig}
\usepackage{amssymb}
\usepackage{soul}
\usepackage[hyphens]{url}
\usepackage{algorithm}
\usepackage{algorithmic}
\usepackage{float}
\newfloat{algorithm}{t}{lop}
\usepackage[nolist]{acronym}%[printonlyused]
\usepackage{lineno,xcolor}
\makeatletter
\newcommand{\specialcell}[1]{\ifmeasuring@#1\else\omit$\displaystyle#1$\ignorespaces\fi}
\makeatother

\usepackage{stfloats}

\begin{document}

%Working title
\title{A Prioritised Traffic Embedding Mechanism enabling a Public Safety Virtual Operator}

\author{\IEEEauthorblockN{
Jonathan van de Belt\IEEEauthorrefmark{1}, Hamed Ahmadi\IEEEauthorrefmark{1}, Linda E. Doyle\IEEEauthorrefmark{1}, Oriol Sallent\IEEEauthorrefmark{2}}
\IEEEauthorblockA{\IEEEauthorrefmark{1}The Centre for Future Networks and Communications \-- CONNECT, Trinity College Dublin} 
%Email: \{vandebej, ahmadih, linda.doyle\} @tcd.ie}
\IEEEauthorblockA{\IEEEauthorrefmark{2}Universitat Polit\'ecnica de Catalunya}
%Email: sallent@tsc.upc.edu}
}
%\thanks{The authors are with CTVR - The Telecommunications Research Centre, Trinity College, The University of Dublin,
%Ireland. %(email: \{finnda, galiotc, pinheirp, vandebej, ahmadih, dasilval\}@tcd.ie).
%L. A. DaSilva is also with Virginia Tech, Arlington, Virginia, USA.}

\maketitle

\begin{abstract}

\ac{ppdr} services can benefit greatly from the availability of mobile broadband communications in disaster and emergency scenarios. While undoubtedly offering full control and reliability, dedicated networks for \ac{ppdr} have resulted in high operating costs and a lack of innovation in comparison to the commercial domain. Driven by the many benefits of broadband communications, \ac{ppdr} operators are increasingly interested in adopting mainstream commercial technologies such as \ac{lte} in favour of expensive, dedicated narrow-band networks. 

In addition, the emergence of virtualization for wireless networks offers a new model for sharing infrastructure between several operators in a flexible and customizable manner. In this context, we propose a virtual \ac{ps} operator that relies on shared infrastructure of commercial \ac{lte} networks to deliver services to its users. We compare several methods of allocating spectrum resources between virtual operators at peak times and examine how this influences differing traffic services. We show that it is possible to provide services to the \ac{ps} users reliably during both normal and emergency operation, and examine the impact on the commercial operators. 

\end{abstract}

%Wireless network virtualization enables multiple virtual wireless networks to coexist on shared physical infrastructure. However, one of the main challenges is the problem of assigning the wireless resources to virtual networks in an efficient manner. Although some work has been done on solving the embedding problem for wireless networks, few solutions are applicable to dynamic networks with changing traffic patterns. In this paper we 
%propose a greedy dynamic embedding algorithm for wireless virtualization. Virtual networks can be re-embedded dynamically using this algorithm, enabling increased resource usage and lower rejection rates. We compare the greedy dynamic algorithm to a static embedding algorithm and also to the adapted dynamic version. We show that the dynamic algorithm provides increased performance to previous methods using simulated traffic. In addition we formulate the embedding problem with multiple priority levels for the dynamic and static case.

\IEEEpeerreviewmaketitle

\begin{acronym}
%% Instructions on use of acronym list:
%%
%% When defining a new acronym include it below using the following format:
%% \acro{<acronym_tag>}[<ACRONYM>]{<Extended form of acronym>}
%%
%% When including an acronym in the main text include as
%% \ac{<acronym_tag>}
%% This will also automatically include the extended form of the acronym on the first occasion it is used.
%%
%% If you wish to force the acronym to be included in extended form, include as
%% \acf{<acronym_tag>}
%%
%% Full details can be found at:
%% ftp://ftp.tex.ac.uk/tex-archive/macros/latex/contrib/acronym/acronym.pdf
%%
%% KEEP THESE IN ALPHABETIC ORDER!

\acro{ppdr}[PPDR]{Public Protection and Disaster Relief}
\acro{ps}[PS]{Public Safety}
\acro{pmr}[PMR]{Professional/Private Mobile Radio}
\acro{sla}[SLA]{service-level agreement}
\acro{mvno}[MVNO]{mobile virtual network operator}
\acro{vo}[VO]{virtual operator}
\acro{inp}[InP]{physical infrastructure provider}
\acro{enodeb}[eNodeB]{enhanced Node B}
\acro{prb}[PRB]{physical resource block}
\acro{vrr}[VRR]{virtual resource request}
%\acro{}[]{}
%\acro{}[]{}
%\acro{}[]{}
%\acro{}[]{}

%\acro{3gpp}[3GPP]{3\textsuperscript{rd} Generation Partnership Program}
%\acro{arq}[ARQ]{Automatic Repeat ReQuest}
%\acro{awgn}[AWGN]{Additive White Gaussian Noise}
%\acro{bcqi}[B-CQI]{Best \ac{cqi}}
%\acro{bler}[BLER]{BLock Error Rate}
%\acro{ca}[CA]{Carrier Aggregation}
%\acro{comp}[CoMP]{Coordinated Multi-Point transmission and reception}
%\acro{cqi}[CQI]{Channel Quality Indicator}
%\acro{csi}[CSI]{Channel State Information}
%\acro{dr}[DR]{Deployment Ratio}
%\acro{dsl}[DSL]{Digital Subscriber Line}
%\acro{edn}[EDN]{Extremely Dense Network}
%\acro{eesm}[EESM]{Exponential Effective \ac{sinr} Mapping}
%\acro{enb}[eNB]{evolved Node Base station} % / Base station?
%\acro{epc}[EPC]{Evolved Packet Core}
%\acro{ffr}[FFR]{Frequency Fractional Reuse}
%\acro{harq}[HARQ]{Hybrid Automatic Repeat reQuest}
%\acro{icic}[ICIC]{Inter Cell Interference Coordination}
%\acro{imt-a}[IMT-A]{International Mobile Telecommunications-Advanced}
%\acro{irc}[IRC]{Interference Rejection Combining}
%\acro{l2s}[L2S]{Link-to-System}
%\acro{ll}[LL]{Link Level}
\acro{lte}[LTE]{Long Term Evolution}
%\acro{lte-a}[LTE-A]{LTE-Advanced}
%\acro{mac}[MAC]{Medium Access Control}
%\acro{mcs}[MCS]{Modulation and Coding Scheme}
%\acro{miesm}[MIESM]{Mutual Information Effective \ac{sinr} Mapping}
%\acro{mimo}[MIMO]{Multiple Input Multiple Output}
%\acro{ml}[ML]{Maximum Likelihood}
%\acro{mmse}[MMSE]{Minimum Mean Squared Error}
%\acro{mrc}[MRC]{Maximum Ratio Combining}
%\acro{mu-mimo}[MU-MIMO]{Multi-User \ac{mimo}}
%\acro{mui}[MUI]{Multi-User Interference}
%\acro{nas}[NAS]{Non-Access Stratum}
%\acro{nl}[NL]{Network Level}
%\acro{oop}[OOP]{Object Oriented Programing}
%\acro{pdcp}[PDCP]{Packet Data Convergence Protocol}
%
%\acro{pf}[PF]{Proportional Fair}
%\acro{phy}[PHY]{Physical layer} % Physical Layer
%\acro{pmi}[PMI]{Precoding Matrix Indicator}
%\acro{pon}[PON]{Passive Optical Network}
%\acro{rb}[RB]{Resource Block}
%\acro{rlc}[RLC]{Radio Link Control}
%\acro{rr}[RR]{Round Robin}
%\acro{rrc}[RRC]{Radio Resource Control}
%\acro{rrh}[RRH]{Remote Radio Head}
%\acro{rrm}[RRM]{Radio Resource Management}
%\acro{sdr}[SDR]{Software Defined Networks}
%\acro{snr}[SNR]{Signal to Noise Ratio}
%\acro{sinr}[SINR]{Signal to Interference and Noise Ratio}
%\acro{siso}[SISO]{Single Input Single Output}
%\acro{sl}[SL]{System Level}
%\acro{sic}[SIC]{Successive Interference Cancellation}
%\acro{su-mimo}[SU-MIMO]{Single-User \ac{mimo}}
%\acro{tbs}[TBS]{Transport Block Size}
%\acro{ts}[TS]{Technical Specification}
%\acro{tti}[TTI]{Transmission Time Interval}
%\acro{ue}[UE]{User Equipment}
%\acro{xgpon}[XG-PON]{10-Gigabit-capable Passive Optical Network}
%\acro{zf}[ZF]{Zero-Forcing}
%
\end{acronym}
\section{Introduction}
\label{sect:introduction}

Mobile broadband communications have great potential to improve the efficiency of \acl{ppdr} operations by enabling applications such as real-time access to high-resolution maps and floor plans, on-field live video transmission from helmet cameras to a central unit, telemedicine, etc. \cite{Report2012} \cite{2CEPT2013}. Driven by this demand, significant changes are taking place in the professional mobile radio industry. While existing \ac{pmr} technologies, such as TETRA, TETRAPOL, DMR, APCO P25, GSM-R and others, have been very successful in delivering business and mission critical voice and narrowband data services in these professional sectors, these technologies are not well suited to support higher data rate applications. In this context, the adoption of mainstream commercial technologies such as \ac{lte}/LTE-Advanced for \ac{ppdr} mission critical communications is gaining strong momentum \cite{TCCA2013}. Indeed, many national and international \ac{ppdr} organizations have endorsed or are considering \ac{lte} as the next generation technology either to augment their existing systems, or to provide a future migration path. The economies of scale of \ac{lte} are expected to bring down the costs of private mobile radio equipment. Establishing common technical standards for the \ac{pmr} and commercial domains offers significant opportunities for exploiting the synergies between these two domains through network and spectrum sharing principles \cite{Ferrus_2013}. Though \ac{lte} technology was originally conceived and designed as a flexible and spectrally efficient mobile broadband technology for the commercial domain, work is in progress within the 3GPP to extend the use of \ac{lte} to critical communication applications and to deliver future services to \ac{ppdr} users \cite{Ferrus_2014}.

In most parts of the world, the prevailing model that has been used for the delivery of mission critical narrowband \ac{ppdr} communications (voice-centric and low data rate services) is based on the use of dedicated  networks; networks deployed for \ac{ppdr} use that are mainly privately built and operated with the specific purpose of serving the communications needs of a limited number of agencies. This delivery model, while undoubtedly offering full control and high availability of communications resources to \ac{ppdr} users, has resulted in a niche market with far less innovation and higher prices for communication equipment in comparison to the commercial domain. 

Given that \ac{ppdr} communications systems are fundamentally funded from constrained public authorities' budgets (i.e., taxpayer's money), \ac{ppdr} communications should strive to achieve the most cost-effective solution.
%as the \ac{ppdr} sector is intimately connected to the public sector of society, 
Without compromising the high control, security and resilience standards required by \ac{ppdr} communications, future broadband \ac{ppdr} data services must be delivered through optimal deployment mechanisms, in a cost-effective manner with a sustainable business model \cite{Ferrus_2014}. 

Acknowledging that while the most suitable model may differ across countries and regions because of differing existing conditions and interests, this paper considers the delivery of mobile broadband services to \ac{ppdr} users over public-access \ac{lte}-based networks. These networks are run by commercial operators and the \ac{ppdr} users share the infrastructure with the rest of commercial traffic. 

Despite some reluctance to use commercial networks from the \ac{ppdr} sector, since it is believed that current commercial networks cannot satisfy many of their requirements (see e.g., \cite{TCCA2013B} \cite{ETSI2010}), there is increasing consensus that these infrastructures undoubtedly have a role to play in critical communications solutions. As commercial broadband networks become an important part of society`s infrastructure, they can enable \ac{ppdr} users to experience the benefits of broadband communications (see e.g., \cite{TCCA2013C} \cite{Erikson2014}). 

In terms of materialization of these principles, \cite{Ferrus_2013} presents a system architecture that enables \ac{ppdr} service access across a number of LTE-based dedicated and commercial network. The system is implemented through a \ac{mvno} model such that no dedicated \ac{lte} networks are deployed for \ac{ppdr}. As an example of practical realization, ASTRID, the national operator of radio communications, paging and the dispatching network for emergency and security services in Belgium, launched a broadband data service called Blue Light Mobile \cite{BlueLightMobile} in Spring 2014. This service allows \ac{ppdr} users to use the commercial 3G networks for data-centric applications. To that end, ASTRID takes on the role of an \ac{mvno} that offers services via the mobile networks of the three national operators in Belgium. 

In turn, sustained on the emergence of virtualization techniques, \cite{vandeBelt2014} proposes a shared physical infrastructure where multiple heterogeneous \acp{vo} coexist concurrently in a much more sophisticated and resource-efficient manner than is the case for current \acp{mvno}. In this model, \acp{vo} are able to request resources in real-time from a \ac{inp} based on detailed requirements, and the virtual embedding problem is dynamically addressed through embedding algorithms that can re-embed existing virtual networks to achieve more efficient resource allocation and satisfy additional resource requests. 

%Initially 2G and 3G access (GSM/EDGE and UMTS/HSPA) are provided, while it is expected to extend the service to LTE.

Given these emerging capabilities, this paper investigates whether a virtual \ac{ps} operator can exist on a virtualized commercial \ac{lte} network. We examine the reliability of such a network for \ac{ps} services, and the benefits to the network owner. 
\section{System Overview}
\label{sect:system_overwiew}

\subsection{Wireless Network Virtualization}

In wireless network virtualization full isolation is required between different virtual networks, so that each virtual network has the illusion of full control over its resources. Virtual networks should not be able to interfere with each other. Link isolation can be achieved by dividing the physical resources into orthagonal dimensions to prevent interference \cite{Park2009}. 

For this work we consider the virtualization of a single \ac{enodeb} in a \ac{lte} network. In this model, a hypervisor is added on top of the physical hardware of the \ac{enodeb} that abstracts the physical \ac{enodeb} resources and manages the resources for several virtual \acp{enodeb} \cite{Zaki2011}. The hypervisor is also responsible for assigning the \ac{lte} spectrum resources (known as \acp{prb}) amongst the virtual networks. Since the \acp{prb} consist of two orthogonal dimensions, frequency and time, full isolation can be ensured between virtual networks. 

On this physical \ac{enodeb} several \acp{vo} can exist, 
%we have one \ac{ppdr} virtual operators and several commercial \acp{vo},
each with their own virtual \acp{enodeb}. For now we assume that sufficient computational resources exist to satisfy the needs of the virtual \acp{enodeb}, and that the computational resources are fully isolated. We are interested only in the \ac{prb} assignment. 

At regular time intervals or rounds, the hypervisor performs the \ac{prb} assignment. We assume that the time between consecutive rounds is such that each of the \acp{prb} have the same average throughput and are equally adequate in the long run. This means the hypervisor can assign any of the \acp{prb} to a virtual network rather than assigning specific resource blocks.

%is only concerned with the number of resources to assign to each virtual operator, rather than dealing with assigning individual resources. 

At any time, \acp{vo} can make \acp{vrr} to the hypervisor on behalf of their users for $r$ \acp{prb}. To ensure that this model applies to both uplink and downlink resource allocation in LTE, the \acp{prb} resources allocated to each user must be contiguous \cite{3GPP2013}. 
% \ac{vrr} requires a number of contiguous \acp{prb}, $r$ \cite{3GPP2013}. 
%Therefore time and frequency resources, $f$ and $t$ respectively, must be requested as a rectangle with an area $f \times t$ so that the \aclp{prb} assigned to each user are contiguous (see figure  \ref{fig:1}). It is up to the \acp{vno} to know how many resources are required for each user and to map this to a number of time and frequency resources.
Therefore, upon receiving a \acl{vrr} for $r$ resources, the hypervisor maps this to a rectangle $f \times t$, with time and frequency dimensions $f$ and $t$ respectively, which ensures that the \aclp{prb} assigned to each user are contiguous.

%this as a rectangle with an area $f \times t$, which are the frequency and time dimensions respectively, so that the \aclp{prb} assigned to each user are contiguous.
%a number of time and frequency resources,

In addition to specifying the number of resources desired, \acp{vrr} also specify a duration, $d$, for the number of rounds that the resources are desired. A priority level, $p$, is also assigned to \acp{vrr}, which allows \acp{vo} to prioritise requests depending on the service type, $s$. Thus each request for a service $s$, \ac{vrr}($s$), has three parameters: $\{p, r, d\}$.

Arriving \acp{vrr} are stored in a buffer until the next resource allocation round. At each round, an attempt is made to allocate resources to all \aclp{vrr} in the buffer, within the constraints of the total resources available, 
$F \times T$, where $F$ and $T$ are the frequency and time dimensions of the substrate. This process is also known as the embedding process. If there are enough free resources to embed a \ac{vrr}, the request is embedded and deemed successful. The virtual operator that made the request can then use the resources to satisfy a user service for the specified number of rounds ($d$). If a \ac{vrr} is not successfully embedded during a round, it remains in the buffer and is embedded in the next round if possible. A \ac{vrr} that is not embedded for a number of rounds, $max\_delay$, is removed from the buffer and the \ac{vrr} is rejected. $max\_delay$ can vary for different services and priority levels. 

%SThus every \ac{vrr} has four parameters associated with it: its priority, duration, and the time and frequency requirements of PRBs.

%For now, we assume that we are only concerned with a single base station, i.e. one geographical location so that the virtual networks do not change location when reassigned.
%The wireless substrate is divided into a number of frequency and time-domain resources, represented as $F$ and $T$. The total number of virtual resources available is $F \times T$.

%This is explained in more detail in \cite{vandeBelt2014}. 
%Differing traffic types can be accommodated through various priority levels.

%Isolation is the main reason for having continuous requests. Also we can tie in the LTE FFD standard here if we want. 

\subsection{Embedding Algorithms}
\label{sect:algorithms} 

We examine several embedding algorithms that the hypervisor could employ to maximize the resource occupancy of the substrate and minimize the number of \acp{vrr} that are rejected. The resource allocation problem is an NP-complete knapsack problem with the additional complexity of priority levels. A complete mathematical formulation of the embedding problem with priority levels is given in our previous work \cite{vandeBelt2014}. 

We investigate static and dynamic embedding algorithms based on the heuristic Karnaugh-map algorithm proposed in \cite{Yang2012}. In this algorithm \acp{vrr} are embedded onto a two-dimensional substrate by finding Karnaugh-map like regions of vacant resources. The algorithm attempts to cluster the \acp{vrr} together, so that the number of contiguous unused resources is as large as possible, which allows additional \aclp{vrr} of varying sizes to be embedded. At each round, the set of \acp{vrr} in the buffer are ordered by priority and decreasing area. Then for each \ac{vrr}, the Karnaugh map approach is used to find a list of free resources with dimensions equal to or greater than the dimensions of the \ac{vrr}. Next the smallest region that can contain the request is chosen. However, within this region it is necessary to find the best corner at which to embed the \ac{vrr}.

The Embedding Density Index (EDI) is used to find the corner that maximizes resource clustering. The EDI measures the number of edges between free and occupied resource blocks. A high EDI reflects a substrate where the   occupied resources are spread out, whereas a substrate in which the occupied resources are grouped together has a low EDI. The effect of embedding the \ac{vrr} at a corner is tested by calculating the resulting EDI. This is done for each of the four corners in the region chosen to contain the request and the corner with the lowest EDI is selected as the embedding location. This ensures that the new virtual service is clustered among the existing virtual services to the greatest degree possible. Further details are provided in \cite{Yang2012}.

In the static case, successful \acp{vrr} receive the same set of resources every round. However with this static embedding scheme it is possible that a virtual network request is rejected due to topology limitations even though there are enough resources available to satisfy it. To overcome this drawback, we consider the dynamic version of the Karnaugh map algorithm, from our previous work \cite{vandeBelt2014}. It dynamically reassigns the existing services every round. The resources can be used much more efficiently and additional \acp{vrr} can be accepted. 

\subsection{Priorities during Normal and Emergency operation}

The hypervisor has two modes of operation: normal operation and emergency operation. During normal operation a virtual resource request from a given service \ac{vrr}($s$) has an associated priority level $p$. When an emergency occurs, the priority levels can be changes on a per-service and per \ac{vo} basis. Thus, for example, during normal operation the Public Safety \ac{vo} may be treated in the same manner as any other commercial operator, but during emergencies, certain services of the Public Safety \ac{vo} can be given higher priority.

\begin{figure*}[!t]
\centering
  \subfloat[All services]{%
    \includegraphics[width=.46\textwidth]{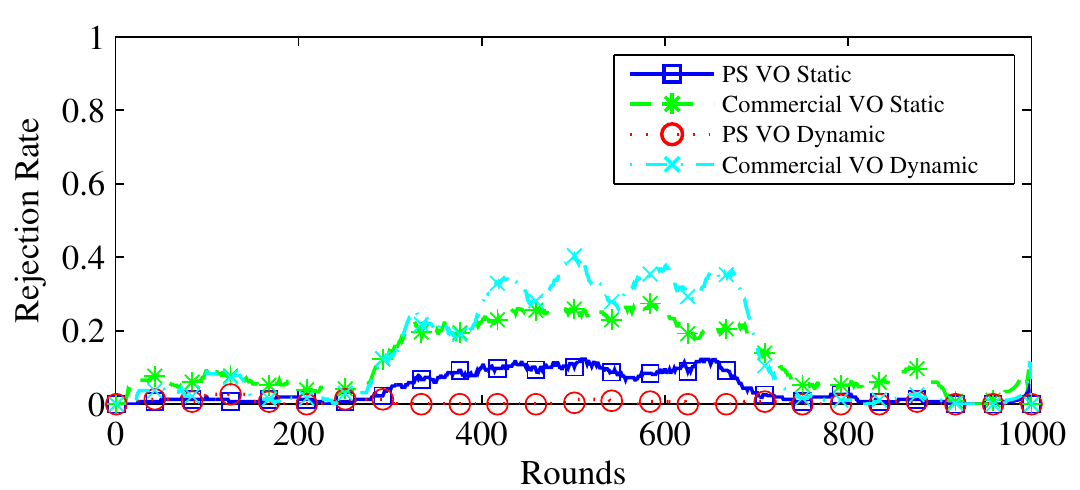}}\hfill
  \subfloat[Voice services]{%
    \includegraphics[width=.46\textwidth]{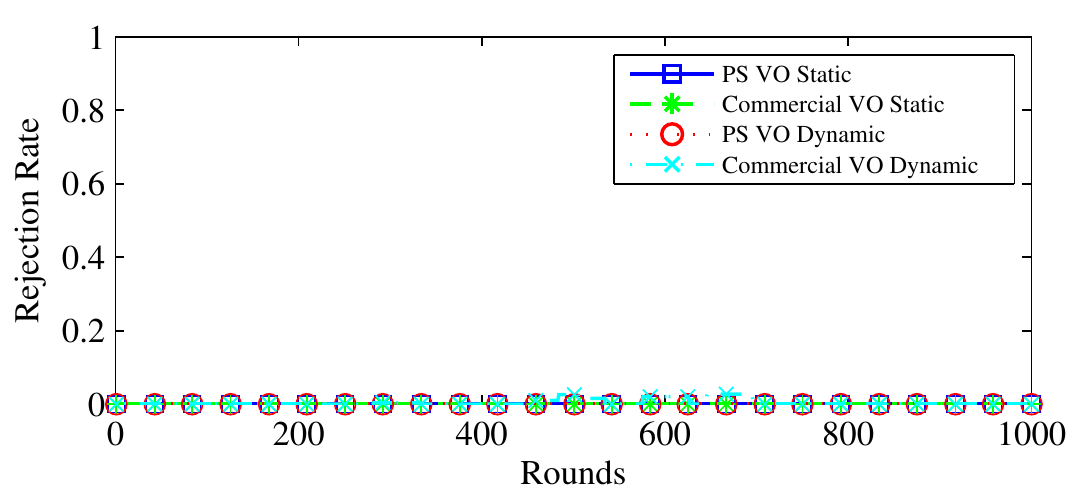}}\hfill
 \subfloat[Video services]{%
    \includegraphics[width=.46\textwidth]{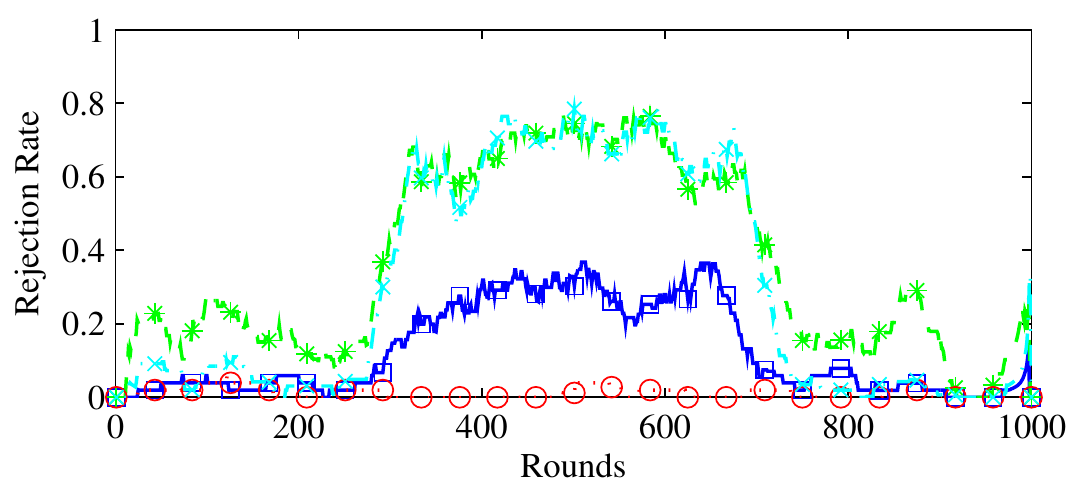}}\hfill
  \subfloat[Messaging services]{%
    \includegraphics[width=.46\textwidth]{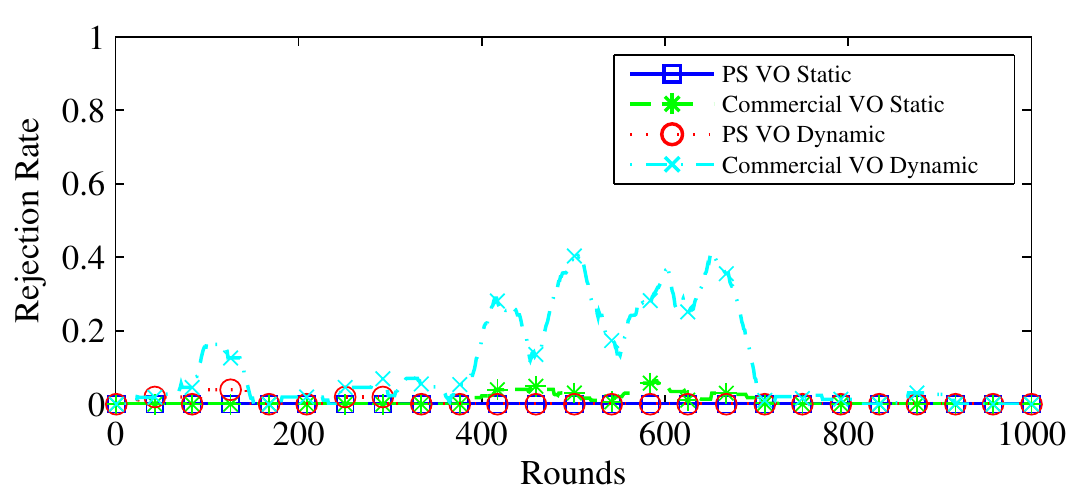}}\hfill
  \caption{Even during an emergency (rounds 300-700), negligible rejection rates for the \acl{ps} services can be achieved for all services (a) using the dynamic embedding algorithm, while for the commercial operator voice services (b) are completely reliable, messaging services (d) are quite reliable, but video services (c) are unreliable.}\label{fig:3}
\end{figure*}

% traffic from all \acp{vo} is considered equal. This means that the \acl{ppdr} \ac{vo} is treated in the same way as any other commercial operator and their \aclp{vrr} are not given special priority. The only priority levels considered by the hypervisor are \acp{vrr} priorities ($p$). 
%
%When an emergency occurs, the public safety operator's \acp{vrr} now receive the highest priority. The commercial \acp{vrr} are embedded after the \ac{ppdr} requests have been satisfied. 
%
%So in essence there are two priority levels: the priority level of \acp{vrr} and the priority level of the \ac{vo} making the request. During normal operation the \ac{vo} priority level is not used, but in emergencies it provides \ac{ppdr} services with exclusive access to the \acp{prb}.

%\acp{vrr} can be made by traffic type, which reflects the priority with which the virtual resources will be allocated. 

\section{Investigated Scenario}
\label{sect:simulation}

The scenario we are investigating is that of a commercial \ac{lte}-based network, where the \acl{inp} has virtualized the mobile resources and \aclp{vo} can request these resources as needed. In our case there is one virtual \ac{ps} network operator and one virtual commercial \ac{vo} that share the physical network.

%These \acp{vo} can have differing market shares and service profiles, based on their business cases.

\subsection{Evaluation Metrics}

The metrics of interest in this scenario are the percentage of \aclp{vrr} that are rejected per service and the percentage of total resources used per service at every round. The rejection rate for a round is calculated as the sum of $r \times d$ of rejected \acp{vrr}, divided by the sum of $r \times d$ of all \acp{vrr}. The first metric shows the importance of reliability from the perspective of the \acp{vo}, while the latter metric provides the \ac{inp} with a method of measuring how efficient its resources are being used. Based on these metrics the \ac{inp} could decide whether additional investment in physical infrastructure is required, and virtual operators can discover whether the \ac{inp} provides a satisfactory service level.

\subsection{Traffic Modelling}
\label{subsec:traffic}

We model three common types of services that each operator desires to fulfil for its users, namely voice traffic, video traffic, and messaging traffic, corresponding to 3GPP's CQI classes 2, 3, and 6 respectively \cite[Table 6.1.7]{3GPPReport2014}. The number of resources, $r$, required for a service is modelled as a uniform distribution $U(min, max)$, and this is then mapped to a permutation of the smallest rectangular area covering at least $r$ \acp{prb}. The duration of each \ac{vrr}, $d$, follows an exponential distribution with an average lifespan of $\mu$ rounds.  

%Based on the 3GPP classes, we assigned the highest priority to voice traffic, medium priority to the video traffic and the lowest priority for the TCP-based traffic. 

%Each \ac{vrr} for a traffic service has several attributes. 
%The number of resources required for any service, $r$, is modelled as a uniform distribution $U(min, max)$, and this is then mapped to a permutation of the smallest rectangular area needed. The duration of each request follows an exponential distribution with an average lifespan of $\mu$ rounds. 

%For voice traffic services the following values are used: $min=1, max=2, \mu=30$. This represents the small number of \acp{prb} that are needed per user, but the duration is quite long in comparison to the other traffic types. The values $min=8, max=25, \mu=10$ apply to the video traffic services, which reflects the high number of \acp{prb} needed, while the duration is of average length. In the case of TCP-based traffic, the number of \acp{prb} needed can fluctuate significantly while the duration is quite short, therefore the values $min=1, max=8, \mu=3$ are set. 

We model the three services as follows:
\begin{align*}
\label{eqn1}
 &\ac{vrr}(voice) &=  & &\{&p, d=30, r \sim U(min=1, max=2)\} \\
&\ac{vrr}(video) &=  & &\{&p, d=10, r \sim U(min=8, max=25)\} \\
 &\ac{vrr}(msg) &=  & &\{&p, d=3, r \sim U(min=1, max=8)\} 
\end{align*}

The maximum delay, $max\_delay$, is set to 1 round for voice service requests, 2 rounds for video services, and 4 rounds for messaging services. The values for these parameters reflect realistic conditions to the best of our knowledge. 

\subsection{Service Priorities}

During normal operation, the \acp{vrr} made by the \ac{ps} and commercial \acp{vo} are considered as having equal importance and priority levels are based only on the service type: the highest priority is for voice requests, next is video and then messaging. When an emergency occurs, all \ac{ps} traffic is given higher priority than commercial traffic, retaining the same service priority levels as before. Regarding commercial traffic, messaging traffic is considered more important during an emergency than video traffic. To this end, higher priority is given to voice, then messaging and eventually video.

\subsection{Simulation setup}

In this scenario the dimensions of the substrate are set to $F = 20, T = 20$. The resource embedding is performed for 1000 rounds. Initially, normal, day-to-day operation is assumed for the hypervisor, however after 300 rounds an emergency occurs. This means that the priority levels of \acp{vrr} change as described above. This emergency situation lasts for 400 rounds, after which the hypervisor switches back into normal operation. 

The number of \acp{vrr} of each service arriving per round is modelled as a Poisson process with the following average rates $\lambda$ requests per round: For normal operation, it takes the values 1.4, 1.4, and 3 for commercial voice, video and messaging respectively, and the values 0.14, 0.14, and 0.3 for \ac{ps} voice, video and messaging. These values offer a realistic approach, since the total percentage of each service type is similar to what is found in the literature and in real systems today \cite{Cisco2014}. Regarding the traffic share between commercial and PS, it assumes that PS represents 10\% of the traffic.
When an emergency occurs, the number of \acp{vrr} increases significantly. The \ac{ps} now requires many more resources and we multiply the values by 5 for all \ac{ps} services. The commercial voice and messaging traffic are also assumed to increase by a factor of 2.5, while we consider that the video traffic remains constant. 

%Each service and During normal day-to-day operation, we assume that almost all of the traffic is for the commercial \acp{vo}. For the commercial traffic, $\lambda$ is set to 1.5 for voice traffic, for video services $\lambda$ is 1.5, while for TCP-based services a value of 1.5 is used. The corresponding values for $\lambda$ for the \ac{ppdr} traffic are 0.1, 0.1 and 0.1 respectively. The percentage of each \ac{vrr} type in our scenario can be seen in figure \ref{fig:2}.

%The number of VRRs of each service arriving per round is modelled as a Poisson process with the following average rates (requests/round): For normal operation, it takes the values 1.4, 1.4, and 3 for commercial voice, video and messaging respectively, and the values 0.14, 0.14, and 0.3 for PS voice, video and messaging. These values offer a realistic approach, since the total percentage of each service type is similar to what is found in the literature and in real systems today [17]. Regarding the traffic share between commercial and PS, it assumes that PS represents 10%.
%When an emergency occurs, the number of VRRs increases significantly. The PS now requires many more resources and we multiply the values by 5 for all PS services. The commercial voice and messaging traffic are also assumed to increase by a factor 2.5.

\section{Performance Results}
\label{sect:performance}

We compare the performance of the embedding algorithms in normal and emergency operation and examine whether virtualized LTE networks can offer adequate service for \ac{ps}. 

The performance achieved in terms of rejection rate per round is depicted in Figure \ref{fig:3}. Regarding the performance of the embedding algorithms, it can be observed in (a) that the dynamic embedding achieves 0\% rejection for all the offered \ac{ps} traffic during the emergency, while the static algorithm rejects about 10\% of the \ac{ps} \acp{vrr}. Since lower priority has been assigned to the commercial traffic during the emergency, higher rejection rates (in the order of 30\% - globally across all commercial services) are observed. Based on these rejection rates, \ac{ps} communication can be achieved reliably over this system.

In more detail, regarding the treatment of the different service flows during the emergency, (b) reflects that all voice traffic from both \ac{ps} and commercial is served. For video service, (c) reveals that the lowest priority given to commercial video traffic during the emergency leads to high rejection rates, in the order of 70\%. Regarding \ac{ps} video traffic, it is observed that the dynamic embedding algorithm is able to provide almost 0\% rejection rate, while the static embedding leads to substantial rejections (above 20\%). Finally, for the messaging service, (d) shows that rejections only occur for commercial traffic and the dynamic embedding algorithm. This is because the dynamic embedding has led to increased allocation of resources to \ac{ps} traffic.

\begin{figure*}[!t]
\centering
  \subfloat[Static embedding VO split]{%
    \includegraphics[width=.46\textwidth]{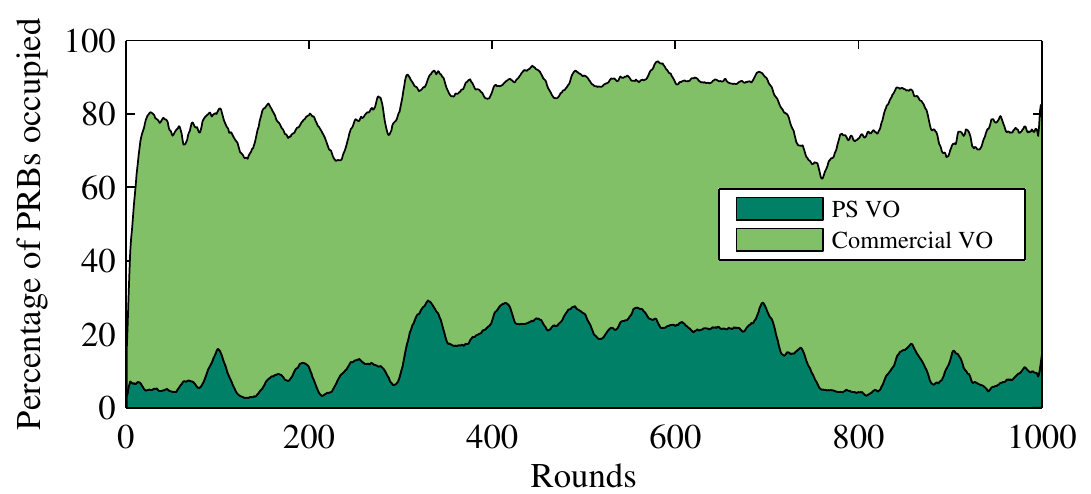}}\hfill
  \subfloat[Dynamic embedding VO split]{%
    \includegraphics[width=.46\textwidth]{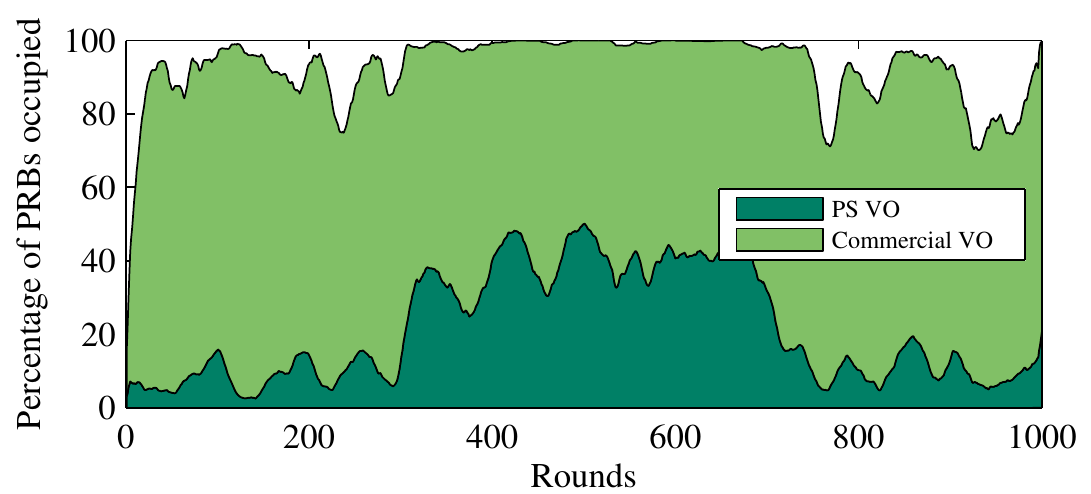}}\hfill
  \subfloat[Static embedding service split]{%
    \includegraphics[width=.46\textwidth]{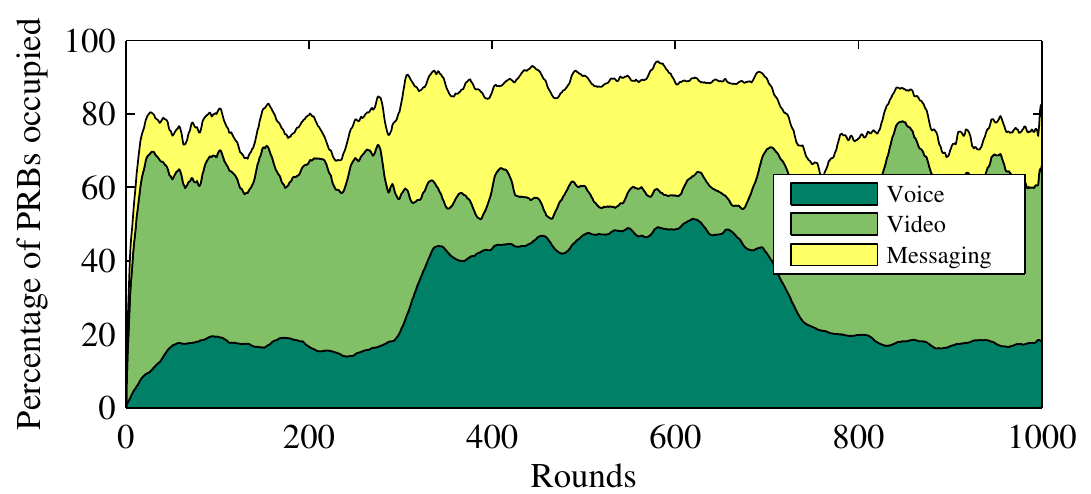}}\hfill
  \subfloat[Dynamic embedding service split]{%
    \includegraphics[width=.46\textwidth]{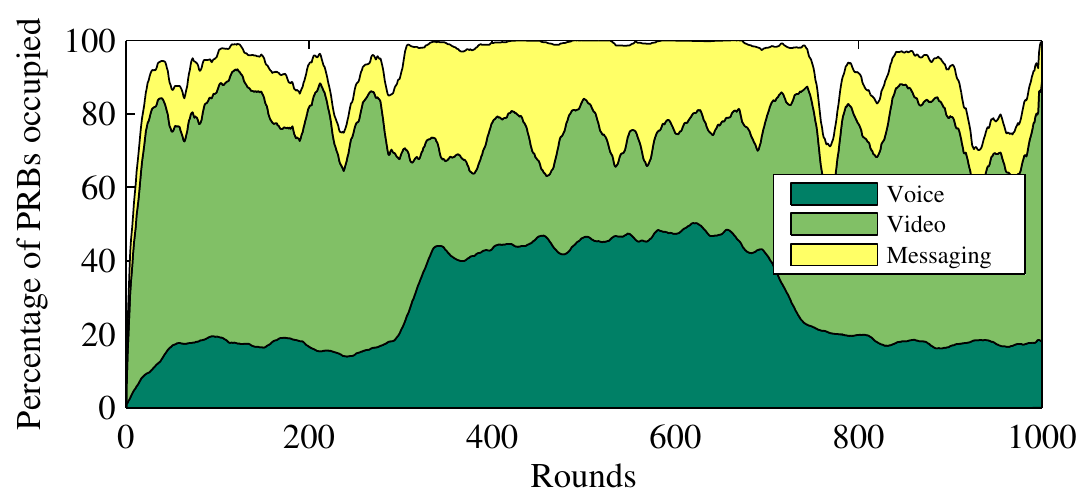}}\hfill
  \caption{Compared to the static embedding (a), the \acl{ps} operator recieves a significantly higher share of total resources when dynamic embedding is used (b). This is mostly due to the low number of video services that are embedding in the static case (c), in comparison to the dynamic case (d).}\label{fig:4}
\end{figure*}

The performance achieved in terms of resources used per round is depicted in Figure \ref{fig:4}. The comparison of (a) and (b) illustrates how the dynamic algorithm is able to exploit almost 100\% of the resources during the emergency, while the static embedding exhibits lower resource exploitation. During normal operation, the resource usage does not reach 100\% because the offered traffic during this period is below the maximum network’s capacity. It can also be observed how the capacity share for the \ac{ps} during the emergency increases (it represents around 40\% of the traffic served during the emergency) compared to the normal operation. This is because of the \ac{ps} traffic increase during the emergency period (i.e., higher requested load) as well as the higher priority assigned to \ac{ps} traffic (i.e., higher served load).
The distribution of the network’s capacity across services is shown in (c) and (d). Both the static and dynamic embedding algorithm devote almost 50\% of the capacity to voice during the emergency. However, the dynamic embedding is able to allocate a substantially larger share of video traffic (about 25\% of the total traffic) compared to the static approach, at the expense of a slight reduction of messaging traffic. In any case, as shown in (d), messaging traffic for the dynamic embedding benefits from the ability of the algorithm to reach almost 100\% resource occupation during the emergency period.

\section{Conclusion}

In this work we proposed a virtual public safety operator that shares a commercial \ac{lte} network with virtual commercial operators. Several embedding algorithms were analysed in terms of the rejection rate and the percentage of resources occupied. We examined the reliability of such a system in normal and emergency situations, and how differing traffic services were affected. We showed that it is possible to provide the public safety operator with reliable communication during emergencies, since almost zero services were rejected. The impact on commercial services was also investigated, and though the performance dropped during emergencies, the most crucial services (voice and messaging) were very reliable. 

%In future work we hope to develop a pricing scheme for this system, taking into account the behaviour of users in emergencies, and investigate whether the cost of such a system is lower than current solutions.

These promising results motivate further studies, such as the formulation of utility functions to capture users’ acceptance of the provided service levels during normal and emergency conditions. This would allow more complete design and parametrization of the embedding algorithm. Additionally, the proposed public safety virtual operator model can be detailed further according to actual/future \ac{lte} architectures and 3GPP-based prioritization and Quality-of-Service control mechanisms.

%These promising results motivate further studies as part of future work. On the one hand, the formulation of utility functions to capture users’ acceptance to the provided service levels during normal and emergency conditions would allow a more accurate solution’s design and parametrization of the embedding algorithm. On the other hand, the proposed public safety virtual operator model can be detailed according to actual/future LTE architectures and 3GPP-based prioritization and Quality-of-Service control mechanisms.

\section*{Acknowledgments}

This work is supported by Science Foundation Ireland under Grants No. 10/CE/I1853, No. 13/RC/2077, and 10/IN.1/I3007 and the Spanish Research Council and FEDER funds under ARCO grant (ref. TEC2010-15198). Special thanks to Dr. Ismael Gomez and Dr. Paul Sutton.

\bibliographystyle{IEEEtran}

\bibliography{public-virtual-operator}

% that's all folks
\end{document}